\documentclass[aps,prb,preprint,showpacs,amsmath,floatfix,superscriptaddress]{revtex4-1}

\usepackage{graphicx}
\usepackage{amsmath,amssymb}
\usepackage{physics}
\usepackage{braket}
\usepackage{siunitx}
\usepackage{xcolor}

\newcommand{\MoS}[1]{MoS\ensuremath{{}_{#1}}}
\newcommand{\GWo}{}
\def\GWo/{GW$_0$}
\newcommand{\GoWo}{}
\def\GoWo/{G$_0$W$_0$}
\newcommand{\scGWo}{}
\def\scGWo/{sc-GW$_0$}

\begin{document}

\title{Solving the Bethe-Salpeter Equation on a Subspace: Approximations and Consequences for Low-dimensional Materials}


\author{Diana Y. Qiu}
\affiliation{Department of Mechanical Engineering and Materials Science, Yale University, New Haven, Connecticut 06520}
\affiliation{Department of Physics, University of California at
Berkeley, California 94720}
\affiliation{Materials Sciences Division, Lawrence Berkeley
National Laboratory, Berkeley, California 94720}

\author{Felipe H. da Jornada}
\affiliation{Department of Materials Science and Engineering, Stanford University, Palo Alto California, 94305}
\affiliation{Department of Physics, University of California at
Berkeley, California 94720}
\affiliation{Materials Sciences Division, Lawrence Berkeley
National Laboratory, Berkeley, California 94720}

\author{Steven G. Louie}
\email{sglouie@berkeley.edu}
\affiliation{Department of Physics, University of California at
Berkeley, California 94720}
\affiliation{Materials Sciences Division, Lawrence Berkeley
National Laboratory, Berkeley, California 94720}

\begin{abstract}
It is well known that the ambient environment can dramatically renormalize the quasiparticle gap and exciton binding energies in low-dimensional materials, but the effect of the environment on the energy splitting of the spin-singlet and spin-triplet exciton states is less understood. A prominent effect is the renormalization of the exciton binding energy and optical strength (and hence the optical spectrum) through additional screening of the direct Coulomb term describing the attractive electron-hole interaction in the kernel of the Bethe-Salpeter equation (BSE). The repulsive exchange interaction responsible for the singlet-triplet slitting, on the other hand, is unscreened within formal many-body perturbation theory. However, Loren Benedict argued that in practical calculations restricted to a subspace of the full Hilbert space, the exchange interaction should be appropriately screened by states outside of the subspace, the so-called $S$ approximation~\cite{Benedict2002}. Here, we systematically explore the accuracy of the $S$ approximation for different confined systems, including a molecule and heterostructures of semiconducting and metallic layered materials. We show that the $S$ approximation is actually exact in the limit of small exciton binding energies (i.e., small direct term) and can be used to significantly accelerate convergence of the exciton energies with respect to the number of empty states, provided that a particular effective screening consistent with the conventional Tamm-Dancoff approximation is employed. We further find that the singlet-triplet splitting in the energy of the excitons is largely unaffected by the external dielectric environment for most quasi-two-dimensional materials.
\end{abstract}


\maketitle

\section{Introduction}

Electron-hole interactions are greatly enhanced in low-dimensional systems as a consequence of both the spatial confinement of the electronic wave functions and the strongly inhomogeneous dielectric environment with a concomitant overall reduction in screening of the Coulomb interaction. For instance, as a consequence of reduced screening in low dimensions, in quasi-one-dimension (quasi-1D), suspended semiconducting carbon nanotubes have strongly bound excitons with binding energies of up to 1 eV~\cite{Spataru2004,Maultzsch2005}, and in quasi-two-dimensions (quasi-2D), suspended monolayers of layered transition metal dichalcogenides (TMDs) have strongly bound excitons with binding energies on the order of 0.5 eV, orders of magnitude larger than their bulk counterparts~\cite{Qiu2013, Mak2013, Chernikov2014, Ugeda2014, Zhu2014, Ye2014,Hill2015}. 
In addition, the electronic and optical properties of quasi-low-dimensional systems are easily tuned by changing the ambient dielectric environment through substrate engineering and encapsulation. In few-layer TMDs, for instance, a graphene substrate or capping layer can renormalize the quasiparticle gap and exciton binding energies by 100-300 meV~\cite{Ugeda2014,Bradley2015,Rasmussen2016}, while encapsulation of few-layer black phosphorus, which displays a weaker intrinsic screening than TMDs, can qualitatively change the optical response from a system with strongly bound excitons to a system whose optical excitations are nearly free-carrier-like~\cite{Li2017,Qiu2017}.
%

Theoretically, excitons can be described with the interacting two-particle Green's function formalism within the \textit{ab initio} GW plus Bethe-Salpeter equation (BSE) or GW-BSE approach~\cite{Strinati1988}. However, two significant challenges arise when applying this formalism to low-dimensional systems. 
First, calculations of low-dimensional systems and nanostructures are typically associated with large unit cells or supercells, so the correlated low-energy neutral excitations often require a large number of band states to be accurately described and are thus computationally demanding. Secondly, while the effect of substrate screening on the quasiparticle (QP) bandgaps and exciton binding energies is basically well established, the effect of the ambient dielectric environment on the energy splitting of the spin-singlet and spin-triplet exciton states has not been as thoroughly explored; and, as we argue here, much care is needed to account for its effect within the \textit{ab initio} GW-BSE approach. For example, obtaining accurate theoretical predictions of the singlet-triplet splitting is essential for describing singlet fission processes (the disossciation of one singlet exciton into two triplet excitons) in some organic photovoltaic materials~\cite{Smith2010,Rangel2016,Refaely2017}, exciton lifetimes~\cite{Spataru2005,Palummo2015}, and exciton dispersion and topology~\cite{Qiu2015}.

The above two theoretical challenges arise as a consequence of the truncation of the Hilbert space for computational efficiency when solving the BSE -- i.e. only a subspace $A$ associated with low-energy bands close to the Fermi energy is explicitly included in solving the BSE. The remaining subspace $B$ of the Hilbert space -- associated with higher energy bands and/or other physical subsystems such as substrates -- are often completely neglected. This approximation comes into two separate parts of the GW-BSE formalism: a) the construction of the electron-hole interaction terms in the BSE kernel, and b) the construction of the basis functions for the exciton wavefunction in solving the BSE. In general, different physics and different cutoffs for the partition into subspaces A and B come into these two parts.  In this work, we focus on part a), which is on the construction of the terms in the BSE kernel. For part a), in Ref.~\onlinecite{Benedict2002}, Benedict argued that, instead of completely neglecting subspace $B$, one could capture its effect by modifying the interaction kernel defined in subspace $A$. The resulting scheme, known as the $S$ approximation, amounts to simply solving the BSE on the subspace $A$, but screening the exchange interaction kernel, $K^x$, in the BSE -- which would otherwise involve only the bare Coulomb interaction -- using a screening scheme that includes only free electron-hole pair excitations from the subspace $B$~\cite{Benedict2002}. 

Benedict \textit{et al} later applied the $S$ approximation to hydrogenated Si clusters~\cite{Benedict2003}, and similar approaches to screening the exchange term in the BSE have been used to obtain better agreement with experiment for various nanocluster systems~\cite{Franceschetti1998,Demchenko2006}, as well as to  study carbon nanotubes on metallic substrates~\cite{Spataru2013}. Recently, Ref.~\cite{Deilmann2019} has explored the effect of applying the S-approximation for benzene on a gold surface, using the RPA dielectric function for gold to screen the exchange term when solving the BSE for an isolated benzene molecule, and found that this form of screening results in a large renormalization of the singlet-triplet splitting of the lowest energy exciton in benzene.
These applications of the $S$ approximation further suggest that, in general, the screening environment might have a large effect on the singlet-triplet splitting in low-dimensional systems and hence on their exciton lifetimes, dispersion, and dynamics, since the exchange interaction kernel $K^x$ is responsible for the splitting in energy between spin-singlet and spin-triplet excitonic states. However, there has been to date no systematic analysis of the accuracy of the $S$ approximation against explicitly full-Hilbert-space converged BSE calculations, nor has there been any comparison of the effect of the substrate on the singlet-triplet splitting in the $S$ approximation against an \textit{explicit} BSE calculation including both substrate and adsorbate. Moreover, the $S$ approximation belongs to a broad class of embedding schemes used to downfold the many-body Hamiltonian across quantum chemistry and condensed matter physics, including in dynamical mean-field theory (DMFT)~\cite{Kotliar2006}, the constrained random phase approximation (constrained RPA)~\cite{Aryasetiawan2006}, self-energy embedding theory (SEET)~\cite{Zgid2017}, and configuration interaction calculations~\cite{Dvorak2019a,Dvorak2019b}, but it is one of the few schemes where the larger Hilbert space remains computationally accessible thus allowing for numerical validation of the embedding procedure.

In this paper, we rederive the $S$ approximation using the L{\"o}wdin partitioning scheme, similarly to Ref.~\onlinecite{Deilmann2019}, and perform a systematic analysis of its accuracy by comparing the solution of the BSE with and without the $S$ approximation as a function of the size of subspace $A$ with respect to the full Hilbert space. We investigate three prototypical systems: (1) an isolated benzene molecule, where subspace $A$ (the low energy orbitals) and $B$ (the high energy orbitals) overlap in real space but not energy space; (2) a monolayer of hexagonal boron nitride (h-BN) encapsulated in bulk \MoS2, where subspace $A$ (h-BN orbitals) and $B$ (\MoS2 orbitals) do not overlap in real space but subspace $B$ is known to screen subspace $A$; and (3) a superlattice of monolayer h-BN and graphene, where again subspaces $A$ (h-BN) and $B$ (graphene) do not overlap in real space but graphene can be expected to strongly screen excitations in h-BN. Tow major results are obtained. (1) We find that, if a carefully designed subspace screening is employed that is consistent with the Tamm-Dancoff approximation (TDA), the $S$ approximation can significantly speed up convergence of the excitation energies with respect to the size of subspace $A$ in molecules in the gas phase where the exchange interaction couples a large number of unoccupied orbitals. We emphasize that, consistent with the usage of the TDA, it is essential to only include electron-hole pairs that are forward-propagating in time -- a subtlety which is either neglected or not made explicit in previous work~\cite{Benedict2002,Spataru2013,Deilmann2019}. (2) We find that both metallic and semiconducting substrates have a negligible effect on singlet-triplet splitting in layered materials, where the wavefunctions of the adsorbate and substrate do not significantly hybridize, and we show that this is true in general for any adsorbate where the electron and hole band wavefunctions can be described by Wannier functions whose extent is smaller than the adsorbate-substrate distance.
\

This article is organized as follows. In section II, we rederive the $S$ approximation. In section III, we describe the details of our implementation of the $S$ approximation. In section IV, we report our results for an isolated benzene molecule. In section V, we report our results for monolayer h-BN encapsulated in bulk \MoS2 and a superlattice of alternating layers of monolayer h-BN and graphene. We conclude in section VI by summarizing our results.

\section{The $S$ Approximation}

Excitons can be described within the two-particle Green's function formalism via the Bethe-Salpeter equation (BSE)~\cite{Strinati1988}
\begin{equation}
\label{eqn:L}
L = L_{0} + L_{0}KL,
\end{equation}where $L$ is the electron-hole correlation function, $L_{0}$ is the noninteracting counterpart describing the free propagation of an electron and a hole (as quasiparticles), and $K$ is an interaction kernel. The BSE is often written, within the Tamm-Dancoff approximation (TDA)~\cite{Strinati1988} and a static approximation for $K$, in terms of an effective eigenvalue problem expressed in a quasiparticle basis,
\begin{equation}
\label{eqn:BSE}
\begin{aligned}
H \psi_X &= \Omega_X \psi_X \\
H &= D + K \\
D_{vc\vb{k},v'c'\vb{k}'} &= \left[\omega - L_0^{-1}\right]_{vc\vb{k},v'c'\vb{k}'} \\
&= \left(\varepsilon_{c\vb{k}} - \varepsilon_{v\vb{k}}\right) \delta_{vv'}\delta_{cc'}\delta_{\vb{k}\vb{k}'},
\end{aligned}
\end{equation}
where $H$ is the effective BSE Hamiltonian; $\psi_X$ and $\Omega_X$ are the associated eigenvectors and eigenvalues, which describe electron-hole neutral excitations of the system; $\varepsilon_n$ are quasiparticle excitation energies;  $v$ ($c$) label valence (conduction) bands; and $\vb{k}$ is a $\vb{k}$ point in the Brillouin zone. For simplicity, the spin index is omitted here.

\begin{figure}
\includegraphics[width=246.0pt]{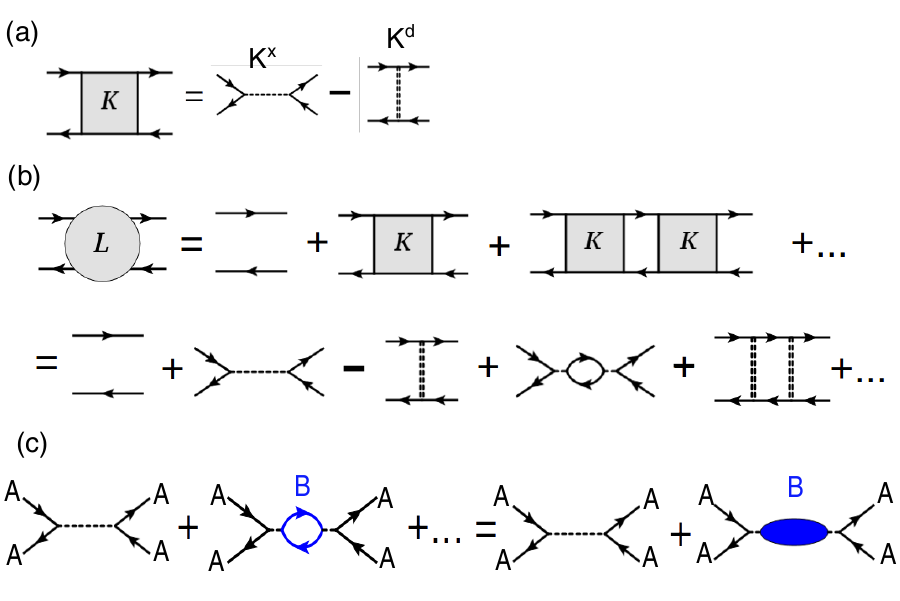}
\caption{\label{fig:diagrams} (a) BSE kernel diagrams consisting of an exchange term ($K^x$) and a direct term $K^d$. (b) Expansion of the electron-hole correlation function ($L$) in the BSE in powers of the BSE kernel. (c) Effective screened exchange interaction kernel $\overline{K}^x_{AA}$, which contains electron-hole bubbles in the $B$ subspace}
\end{figure}

Rigorous conserving approximations for the kernel $K$ can be obtained by expressing it in terms of the functional derivative of the single-particle self-energy $\Sigma$ with respect to the single-particle Green's function $G$, $K=\frac{\delta \Sigma}{\delta G}$. Within the commonly employed GW approximation for $\Sigma$ and neglecting the variation of the dynamically screened interaction when computing the functional derivative~\cite{Strinati1988}, $K$ can be expressed as the sum of an attractive direct term ($K^{d}$) involving the screened Coulomb interaction and a repulsive exchange term ($K^{x}$) involving the bare Coulomb interaction. The exchange term, $K^{x}$, is solely responsible for the splitting of spin-singlet and spin-triplet excitations, if spin-orbit interaction is neglected. The two kernel terms are shown diagrammatically in Fig.~\ref{fig:diagrams}a, and Fig.~\ref{fig:diagrams}b shows the expansion of the electron-hole correlation in Eq.~\ref{eqn:BSE} in powers of $K^x$ and $K^d$. An important property that the interaction kernel $K$ must have in  Eq.~\ref{eqn:BSE} is that it must be a \textit{proper interaction}~\cite{Fetter2003}; i.e., no component of $K$ can be written in terms of two interaction kernels $K^a$ and $K^b$ connected by a noninteracting correlation function, $K\ne K^a L_0 K^b$. This property is automatically satisfied when the kernel is computed in terms of the functional derivative of a conserving electronic self-energy. Following this analysis, one should indeed use the bare Coulomb interaction in $K^x$ instead of the screened counterpart $K^W=K^x+K^x L_0 K^x + \cdots$, as the latter is improper.

While it is straightforward to justify the usage of the the bare interaction in $K^x$ through the aforementioned analytic arguments, the situation is more nuanced in numerical calculations when the BSE is solved in a subspace of the total Hilbert space.
%
%
Motivated by Benedict~\cite{Benedict2003}, we formally separate the Hilbert space into two subspaces. Subspace $A$ -- the one that we wish to explicitly solve for -- may include (i) only a small number of quasiparticle states necessary to describe correlated neutral excitations, and/or (ii) a subset of the physical system of interest, e.g., a molecule or a quasi-two-dimensional (quasi-2D) material but not the substrate supporting it. We denote the remainder of the Hilbert space by subspace $B$.

We write the full BSE effective Hamiltonian in a block form in terms of the two subspaces as
\begin{equation}
H^{\mathrm{BSE}}=\left( 
\begin{matrix}
H_{AA} & H_{AB} \\
H_{BA} & H_{BB}
\end{matrix}
\right),
\end{equation}
where the non-interacting electron-hole correlation function is diagonal in the quasiparticle basis, so any terms involving $L_{0}^{AB}$ or $L_{0}^{BA}$ (and $D^{AB}$ or  $D^{BA}$ in Eq.~\ref{eqn:BSE}) must be zero. We emphasize that this is not an approximation, unlike what was originally stated in the proposal of the $S$ approximation~\cite{Benedict2002}. The form of the energy-dependent effective Hamiltonian projected onto subspace A is described by L{\"o}wdin partitioning~\cite{Lowdin1982} as
\begin{equation}
\label{eqn:Heff1}
\begin{aligned}
    H_\mathrm{eff}(\omega) &= H_{AA}+H_{AB} \overline{L}_B(\omega) H_{BA}\\
    \overline{L}_B(\omega) &\equiv \left(\omega-H_{BB}\right)^{-1},
\end{aligned}
\end{equation}
where $\overline{L}_B(\omega)$ is the interacting electron-hole correlation function for the subspace $B$ alone, i.e., without any cross term with subspace $A$.

The effective Hamiltonian $H_\mathrm{eff}$ in $A$ cannot be obtained without explicit knowledge of $B$, so it is typically approximated by truncating the full Hamiltonian to the $A$ subspace ($H_\mathrm{eff}\approx H_{AA}$). Calculations are then converged with respect to the size of subspace $A$. This approach is reasonable as long as the eigenenergies of $H_{AA}$ are far from those of $H_{BB}$ at convergence, and the coupling of the subspaces $H_{AB}$ and $H_{BA}$ are small. Additionally, even though there are formalisms that solve the BSE without explicitly including unoccupied states~\cite{Rocca2010,Rocca2012}, the Hilbert space is still often separated into subspaces $A$ and $B$, for instance, through the definition of the projection operator or by not explicitly including the supporting substrate in the corresponding Hamiltonian.

An intermediate approximation, which is expected to be more accurate than setting $H_\mathrm{eff}\approx H_{AA}$, is to include only the exchange contribution to the interaction kernel in the coupling terms other than $K_{AA}$. This is physically motivated by the fact that, unlike $K^x$, the direct term $K^d$ depends on the spatial overlap of electron and hole states between initial and final states. Hence, $K^d_{AB}$ and $K^d_{BA}$ are expected to vanish when subspaces $A$ and $B$ are separated in space, such as when subspace $A$ contains the states for a molecule and subspace $B$ the states for the substrate supporting it. Within this approximation, we obtain
\begin{equation}
\label{eqn:Heff2}
\begin{aligned}
	H_\mathrm{eff}(\omega) &=  D_A + K^d_{AA} + \overline{K}^x_{AA}(\omega) \\
    \overline{K}^x_{AA}(\omega) &= K^x_{AA} + K^x_{AB} \overline{L}_B(\omega) K^x_{BA}(\omega)\\
    \overline{L}_B(\omega) &=
    L_B^0(\omega) + L_B^0(\omega) K_{BB} \overline{L}_B(\omega).
\end{aligned}
\end{equation}

We identify  $\overline{L}_B$ in Eq.~\ref{eqn:Heff2} as the electron-hole polarization propagator for subspace $B$ evaluated within the ring approximation~\cite{Fetter2003}, so that $\overline{K}^x_{AA}$ is simply the exchange contribution to the interaction kernel but with the bare Coulomb interaction modified to include screening arising from positive-energy electron-hole pair excitations from subspace $B$. The effective Hamiltonian in Eq.~\ref{eqn:Heff2} is equivalent to solving for the electron-hole correlation function in the subspace $A$ but using a modified screened exchange interaction kernel, which incorporates the coupling of subspace $A$ with subspace $B$. The effective interaction kernel $\overline{K}^x_{AA}$ is shown diagramatically in Fig.~\ref{fig:diagrams}~(c). We note that in the limit of small exciton binding energies or vanishing direct interaction, Eq.~\ref{eqn:Heff2} is exact, as the modified interaction kernel $\overline{K}^x_{AA}$ is equivalent to the downfolding within L{\"o}wdin partitioning~\cite{Lowdin1982}. 

We note that the approximation obtained in Eq.~\ref{eqn:Heff2} is equivalent to the $S$ approximation obtained by Benedict~\cite{Benedict2002} but with an important difference: consistent with the use of the Tamm-Dancoff approximation in the BSE, the screening from subspace $B$ that is used in $\overline{K}^x_{AA}$ only contains positive-energy electron-hole pair excitations as opposed to positive and negative excitations. This is an often overlook detail, which can cause an overscreening within the $S$ approximation if neglected.

Finally, we emphasize that, even if the interaction kernel $K^d$ and $K^x$ are strictly static, the effective interaction kernel $\overline{K}^x_{AA}(\omega)$ picks up a dynamic dependence due to the downfolding procedure. Correspondingly, the screened Coulomb interaction constructed from states from outside the subspace $A$ should rigorously be self-consistently evaluated at the excitation energy $\omega=\Omega_X$ of the state $X$ that one is interested in solving for. However, we drop the dynamical dependence in $\overline{K}^x_{AA}(\omega)$ in this work. This is partially justified by the fact that, when subspace $B$ is separated enough in energy, $\overline{K}^x_{AA}(\omega)$ will be a  smooth function up to an energy scale of the order of the smallest quasiparticle band-to-toband transition energy difference $D_{vc\vb{k},v'c'\vb{k'}}^{B}$ in the $B$ subspace. This is typically much larger than the excitation energy of the low-energy excitons we are interested in when we use the $S$ approximation to accelerate the convergence with respect to the number of  bands included in the $A$ subspace. Yet, this is not strictly the case when the $B$ subspace overlaps in energy with the lowest exciton excitation energy, as in the case when subspace $B$ is a substrate. The static approximation to $\overline{K}^x_{AA}(\omega)$ in this case may cause an additional error, although our calculations show that this error is small, as we show in the next sections.

In the next section, we describe how to implement the $S$  approximation within an \textit{ab initio} GW-BSE framework.

\section{Numeric Implementation of the $S$ Approximation}

In this section, we detail the modifications in the BerkeleyGW software package~\cite{Hybertsen1987,Rohlfing2000,Deslippe2012} to implement the $S$ approximation. For all systems, we solve for the electron-hole excitations within the \textit{ab initio} GW plus Bethe Salpeter equation (GW-BSE) approach in the Tamm-Dancoff approximation (TDA). We modify the kernel of the BSE to implement the $S$ approximation by screening the exchange matrix elements in the following way:
\begin{equation}
\begin{split}
&\langle vc\mathbf{k}|K^x|v'c'\mathbf{k}'\rangle = \\ &\sum_{\mathbf{GG}'} M_{vc\mathbf{k}}^*(\mathbf{Q},\mathbf{G})\overline{W}_{\mathbf{GG}'}(\omega=0,\mathbf{Q})M_{v'c'\mathbf{k'}}(\mathbf{Q},\mathbf{G}').
\end{split}
\end{equation}
Here, the BSE matrix elements are written in the basis of electron-hole pairs $vc$ at a k-point $\mathbf{k}$. On the right-hand side, $M_{vc\mathbf{k}}(\mathbf{Q},\mathbf{G})=\langle v\mathbf{k}+\mathbf{Q} |e^{i(\mathbf{Q}+\mathbf{G})\cdot \mathbf{r}}|c\mathbf{k}\rangle$, where $\mathbf{G}$ is a reciprocal lattice vector, and $\mathbf{Q}$ is the center-of-mass momentum of the electron-hole pair (for simplicity, the numerical examples we show will only consider excitations where $\mathbf{Q}=0$). $\overline{W}_{\mathbf{GG}'}(\omega=0,\mathbf{Q})$ is the screened Coulomb interaction in the static limit. In the absence of the $S$ approximation, $\overline{W}$ is the usual bare Coulomb interaction,
$\overline{W}_{\mathbf{GG}'}(\mathbf{Q}) = v(\mathbf{Q}+\mathbf{G})\,\delta_{\mathbf{GG'}}$,
where $v(\mathbf{Q}+\mathbf{G})$ are the Fourier components of the Coulomb interaction. Since we solve the BSE within the TDA, solutions with positive and negative excitation energies do not mix, and hence, the virtual electron-hole pairs introduced by the exchange diagram in the BSE also do not couple positive and negative excitation energies. Consequently, within the $S$ approximation, $\overline{W}$ is screened by the polarizability of the $B$ subspace created only by positive electron-hole excitations, which, at $\omega=0$, is equal to half of the non-interacting polarizability that is typically computed within the random-phase approximation (RPA). That is,
\begin{equation}
\overline{W}_{\mathbf{GG}'}(\omega=0,\mathbf{Q})=\overline{\epsilon}^{-1}_{\mathbf{GG}'}(\omega=0,\mathbf{Q})v(\mathbf{Q}+\mathbf{G}),
\end{equation}
where $\overline{\epsilon}$ is a dielectric matrix of the form
\begin{equation}
\label{eqn:epsilon}
\overline{\epsilon}_{\mathbf{GG}'}(\omega=0,\mathbf{Q})=\delta_{\mathbf{GG}'}-v(\mathbf{Q}+\mathbf{G})\frac{1}{2}\overline{\chi}^0_{\mathbf{GG}'}(\omega=0,\mathbf{Q}).
\end{equation}
The factor of $\frac{1}{2}$ removes the electron hole-pairs that are backward propagating in time, and $\overline{\chi}^0$ is the static, non-interacting RPA polarizablity due to all electron-hole pairs not in subspace $A$,
\begin{align}
\overline{\chi}^0_{\mathbf{GG}'}(\mathbf{Q})=
\sum_{n}^{\mathrm{occ}}\sum_{n'}^{\mathrm{unocc}} \sum_{\mathbf{k}}\frac{M_{nn'\mathbf{k}}^*(\mathbf{Q},\mathbf{G})M_{nn'\mathbf{k}}(\mathbf{Q},\mathbf{G}')}{E_{n\mathbf{k}+\mathbf{Q}}-E_{n'\mathbf{k}}} \notag
\\ -\left[\chi^{AA}\right]^0_{\mathbf{GG}'}(\mathbf{Q}),\label{eqn:chibar}
\end{align}
while $\left[\chi^{AA}\right]^0$ is the non-interacting RPA polarizability formed by transitions from occupied to unoccupied states both belonging to the $A$ subspace,
\begin{equation}
\label{eqn:chiAA}
\begin{split}
\left[\chi^{AA}\right]^0_{\mathbf{GG}'}(\mathbf{Q})=
    \sum_{n\in A}^{\mathrm{occ}}\sum_{n'\in A}^{\mathrm{unocc}} \sum_{\mathbf{k}}\frac{M_{nn'\mathbf{k}}^*(\mathbf{Q},\mathbf{G})M_{nn'\mathbf{k}}(\mathbf{Q},\mathbf{G}')}{E_{n\mathbf{k}+\mathbf{Q}}-E_{n'\mathbf{k}}}.
\end{split}
\end{equation}
In Eqs.~\ref{eqn:chibar} and~\ref{eqn:chiAA}, we have dropped the $\omega=0$ variable for compactness.


\section{Benzene Molecule}


In this section, we calculate the neutral excitations of a single benzene molecule in a 16~\AA{}~$\times$~16~\AA{}~$\times$~16~\AA{} supercell. First, we use density-functional theory (DFT)~\cite{Hohenberg1964,Kohn1965} in the local density approximation (LDA)~\cite{Perdew1981}, as implemented in Quantum ESPRESSO~\cite{Giannozzi2009}, to obtain a mean-field starting point for our GW and GW-BSE calculation. Plane-wave components up to 80~Ry are included in the wavefunction. We then performed a GW-BSE calculation with the BerkeleyGW codee~\cite{Deslippe2012,Hybertsen1986,Rohlfing2000}. In the GW calculation, the dynamical screening effects were accounted for within the Hybertsen-Louie generalized plasmon-pole (HL-GPP) model~\cite{Hybertsen1986}. The dielectric matrix includes plane-wave components up to 20 Ry, and the summation over unoccupied states included 2000 states. The static remainder technique was used to speed up convergence with respect to the number of empty states~\cite{Deslippe2013}. We truncate the Coulomb potential in a sphere with a radius of 8\AA{} to prevent spurious interactions between periodic images of the supercell. For these parameters, we find a HOMO-LUMO gap of 10.4~eV. For the BSE calculation, we include plane-wave components up to 6~Ry in the dielectric matrix. Then, we solve the BSE with and without the $S$ approximation and examine the convergence of the singlet and triplet excitation energies with respect to the number of unoccupied bands explicitly included in the BSE Hamiltonian. The number of occupied states included in the BSE Hamiltonian is fixed at 10.

\begin{figure}
\includegraphics[width=246.0pt]{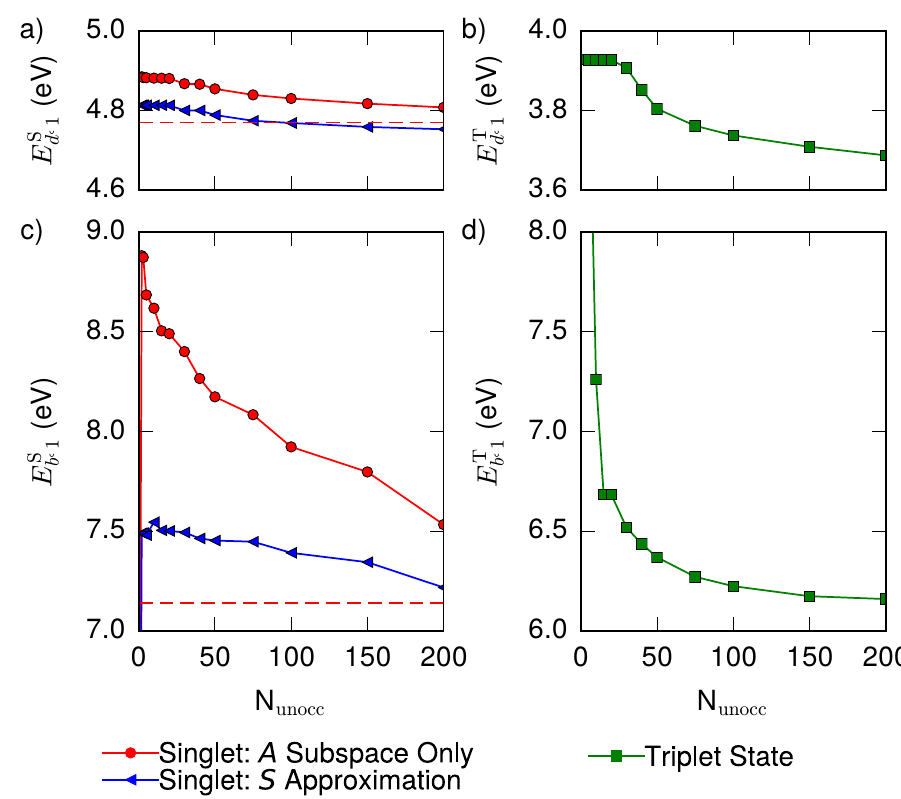}
\caption{\label{fig:benzene} Convergence of exciton excitation energies for a benzene molecule with respect to the number of unoccupied bands (N$_{\mathrm{unocc}}$) used to diagonalize the BSE Hamiltonian of (a) the singlet excitation energy for the lowest energy dark exciton $E_{d,1}^{\mathrm{S}}$ (which is dark due to a separate selection rule), (b) the excitation energy for the triplet state (spin partner to the state in (a)) associated with the lowest energy dark exciton $E_{d,1}^{\mathrm{T}}$, (c) the singlet excitation energy for the lowest energy bright exciton $E_{b,1}^{\mathrm{S}}$, and (d) the excitation energy for the triplet state (spin partner to the state in (c)) associated with the lowest energy bright exciton $E_{b,1}^{\mathrm{T}}$. Red circles give singlet excitation energies calculated in the $A$ subspace only. Blue triangles are singlet excitation energies calculated with the $S$ approximation. The red dashed line is the linear extrapolation of the red circles to $1/\mathrm{N}_{\mathrm{unocc}}\rightarrow 0$. The green squares show the convergence of the triplet state, which does not include any exchange contribution.
}
\end{figure}

We focus on two low-energy exciton states in analyzing the results of the BSE: (1) the lowest-energy excited state, which we label E$_{d,1}$, whose singlet exciton is dark for polarizations of light in the plane of the benzene ring perpendicular to the carbon-carbon bond ; and (2) the lowest-energy exciton state for which the singlet exciton is bright when the polarization of the external electric field is in the plane of the benzene ring perpendicular to the carbon-carbon bond, which we label E$_{b,1}$. To analyze the effect of the exchange interaction, we compare the convergence of the spin-singlet exciton E$^S_{d,1}$ (E$^S_{b,1}$) for state "d"(state "b"), which has a contribution from the exchange interaction, with the convergence of the spin-triplet exciton E$^T_{d,1}$ (E$^T_{b,1}$) for state "d"(state "b"), which has no contribution from the exchange interaction. We note here that both "d" and "b" states can either be spin singlet or spin triplet. Only the "b" spin singlet state is optically active.  For the "d" states, the spin singlet state is optically inactive due to the mirror plane symmetry of the molecular orbitals composing the exciton. The convergence of the singlet states and triplet states with respect to the number of unoccupied states and without the $S$ approximation is shown respectively by the red circles and green squares in Fig.~\ref{fig:benzene}, respectively. We see that the excitation energy of the E$_{d,1}$ singlet and triplet states converge rapidly with respect to the number of unoccupied bands, while the excitation energy of the E$_{b,1}$ singlet converges very slowly, requiring more than 200 bands to converge the error in the excitation energy to less than 100~meV. Including dynamical effects in the screening changes the eigenvalues by less than 50 meV. The triplet state associated with the E$_{b,1}$ singlet state, however, converges much more rapidly with the number of unoccupied bands, suggesting that the slow convergence can be primarily attributed to the exchange interaction, making the E$_{b,1}$ exciton an ideal test case of the $S$ approximation.

\begin{figure}
\includegraphics[width=246.0pt]{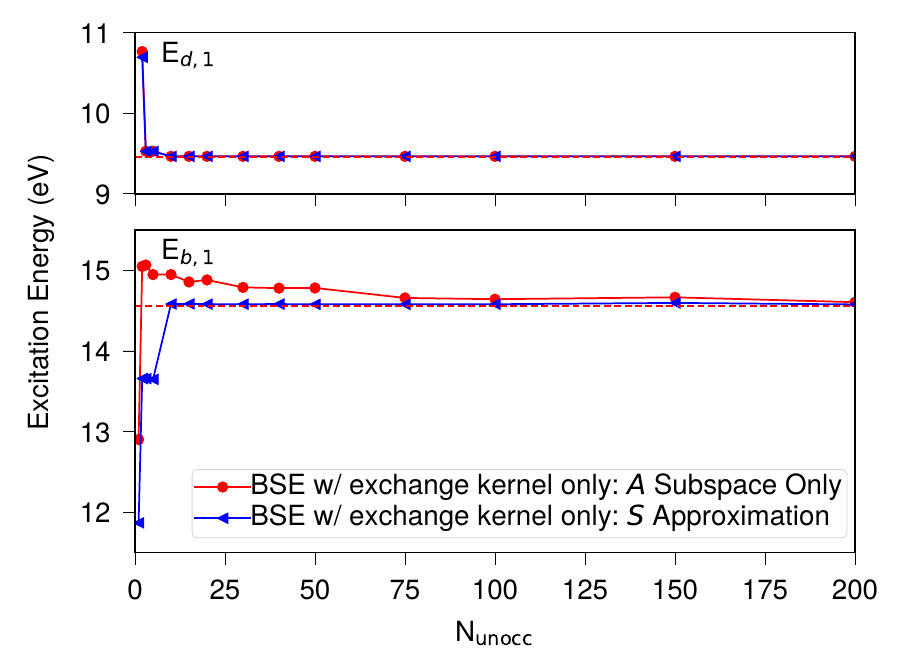}
\caption{\label{fig:benzene exchange} Convergence of singlet excitation energy in a benzene molecule for $E_{d,1}$ (top panel) and $E_{b,1}$ (bottom panel) with respect to the number of unoccupied bands (N$_{\mathrm{unocc}}$) used to diagonalize the BSE Hamiltonian with only the exchange interaction included in the interaction kernel (i.e. $K^d=0)$. Red circles are excitation energies in the restricted subspace only. Blue triangles are excitation energies calculated with the $S$ approximation. The dashed line is the linear extrapolation of the red circles to $1/{\mathrm{N}_{\mathrm{unocc}}}\rightarrow 0$.}
\end{figure}

Next, we apply the $S$ approximation to the benzene molecule by screening the exchange interaction. The results are shown by the blue triangles in Fig.~\ref{fig:benzene}. We see that the $S$ approximation does indeed speed-up convergence with respect to the number of unoccupied states. To understand where this improvement is coming from, we then solve the BSE with the direct Kernel term, which is responsible for the finite binding energy of bound excitons, set to zero -- the limit where the $S$ approximation becomes exact. The convergence of the BSE with only the exchange interaction is shown in Fig.~\ref{fig:benzene exchange}. In the restricted subspace, the convergence of the eigenvalues for E$_{b,1}$ is slow, but when the exchange is screened within the $S$ approximation, the eigenvalue immediately jumps to the converged value after a small number of unoccupied states have been included. These results suggest that the $S$ approximation can be used to moderately speed up convergence of the excitation energies when the exchange interaction couples a large number of unoccupied orbitals.

\section{Subspace Partitioning in Low-dimensional Materials on a Substrate}

\subsection{\label{sec:mos2_bn}Boron Nitride encapsulated in \MoS2}

\begin{figure}
\includegraphics[width=246.0pt]{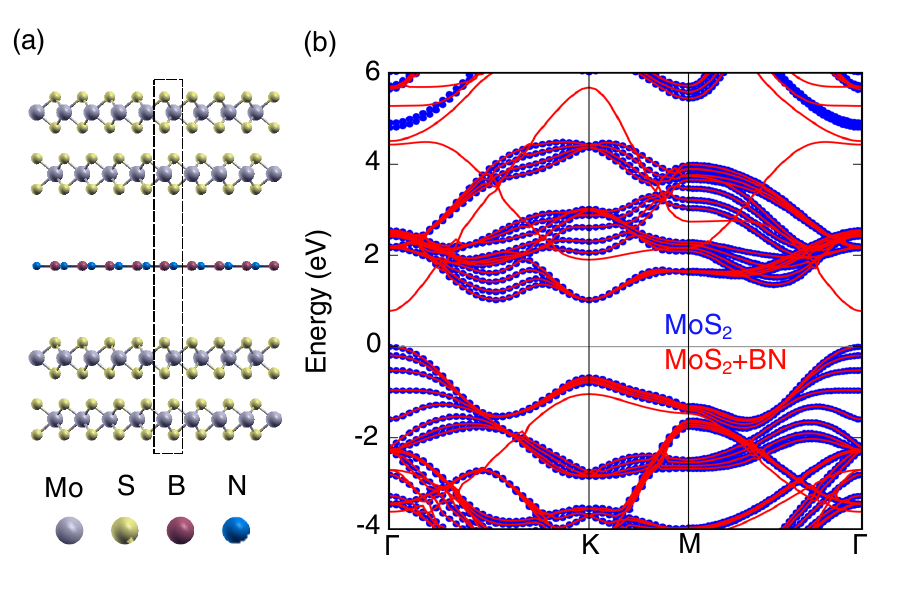}
\caption{\label{fig:mos2_bn} (a) Illustration of supercell with strained h-BN encapsulated in \MoS2. The dashed rectangle indicates the boundaries of the supercell used in the calculation, which is periodic in all directions}. (b) Bandstructure at LDA level of the \MoS2 and h-BN supercell (red lines) superimposed on the bandstructure of an isolated \MoS2 slab (blue circles).
\end{figure}

In this section, we study how the singlet-triplet splitting may be affected by the presence of substrates and whether the $S$ approximation can accurately describe this effect. We first consider the case of a monolayer of h-BN encapsulated between bulk-like slabs (4 layers) of \MoS2. In order to provide a computationally feasible test of the $S$ approximation, we setup an artificial system where the BN is strained by 20\% in order to match the lattice of \MoS2. The \MoS2 and BN slabs are separated by 6\AA{} to prevent hybridization of the wave functions. The supercell setup is shown in Fig.~\ref{fig:mos2_bn}a. In Fig.~\ref{fig:mos2_bn}b, we compare the bandstructure of the supercell with the bandstructure of the \MoS2 slab alone. The overlap of the two bandstructures indicates that there is minimal hybridization of the wavefunctions near the Fermi level. We emphasize that this supercell setup is not meant to describe realistic experimental conditions but rather to provide a computationally tractable test case, where the wave functions of subspaces $A$ and $B$ do not hybridize but the screening environment is dramatically changed from quasi-2D (for a freestanding atomic layer of monolayer h-BN) to bulk-like (for h-BN encapsulated in \MoS2). Here, we focus on the effect of screening from \MoS2 on excitations in h-BN because \MoS2 is more polarizable and h-BN has a large singlet-triplet splitting on the order of 100~meV compared to 20~meV in \MoS2, so any screening effect on the exchange interaction is expected to be more pronounced in h-BN.

In solving the BSE, the dielectric matrix is calculated on a 24x24x1 k-point grid with plane-wave components up to 5~Ry included in the dielectric matrix and band states with energies up to 5 Ry included in the sum over unoccupied states. We calculate two dielectric matrices: one that includes the full supercell of monolayer h-BN encapsulated in a slab of \MoS2 ($\epsilon_{\mathrm{BN+\MoS2}}$) and another that includes only the \MoS2 slab ($\epsilon_{\mathrm{\MoS2}}$). Then, we solve the BSE for three cases: (1) the conventional treatment considering a freestanding monolayer of h-BN, where the direct kernel $K^d$ is screened by the full dielectric matrix $\epsilon_{\mathrm{BN+\MoS2}}$ and the exchange kernel $K^x$ is unscreened; (2) the $S$ approximation case, where $K^d$ is screened by $\epsilon_{\mathrm{BN+\MoS2}}$ and $K^x$ is screened by $\epsilon_{\mathrm{\MoS2}}$; and (3) the explicit BSE calculation on the full supercell containing both h-BN and \MoS2. We compare the singlet-triplet splitting of the lowest-energy exciton in h-BN for the three cases described above. In case (3), when the BSE is solved for the full system including \MoS2 and h-BN, the calculation includes many excitons arising from \MoS2. To identify the spin-singlet and spin-triplet excitons arising primarily from h-BN, we calculated the charge density of the electron contribution to each exciton state when the hole is fixed at the center of the h-BN slab. We identify excitons with maximum electron charge density lies 3~\AA{} of the center of the h-BN slab as excitons in h-BN.

\begin{table}
\caption{\label{tab:stsplit} Singlet-triplet splitting for the lowest-energy exciton in monolayer h-BN encapsulated in \MoS2 (\MoS2 + h-BN) and in a superlattice with alternating layers of monolayer h-BN and graphene (graphene + h-BN) calculated in three ways: 1) with the basis of the BSE matrix restricted to the subspace of h-BN ($A$ subspace), 2) with the $S$ approximation, and 3) with the full Hamiltonian including states from both h-BN and \MoS2 or graphene.
}

\begin{ruledtabular}
\begin{tabular*}{\textwidth}{lccc}
\hline
~ & \multicolumn{3}{c}{Singlet-Triplet Splitting (eV)} \\
\cline{2-4} \\
~ & $A$ subspace & $S$ Approx. & Full Hamiltonian \\

\hline

\MoS2 + h-BN & 0.07 & 0.07 & 0.06 \\ 
graphene + h-BN & 0.1 & 0.1 & 0.09 \\
\hline
\end{tabular*}
\end{ruledtabular}
\end{table}

The singlet-triplet splitting of the lowest-energy h-BN exciton for the three cases is shown in the first row of Table~\ref{tab:stsplit}. We see that encapsulation causes a small reduction in the singlet-triplet splitting. Explicitly including the \MoS2 states in the BSE Hamiltonian reduces the singlet-triplet splitting by 5~meV or about 7\%. The $S$-approximation approximation gives the same result as the freestanding h-BN. Here, the small error in the $S$ approximation, compared to the full BSE Hamiltonian, arises from neglecting wavefunction hybridization between subspaces $A$ and $B$ and the assumption that the screening from subspace $B$ is static. We also perform an additional calculation at a reduced distance of 3.5\AA{} to isolate the effect of substrate distance. At the reduced distance, hybridization effects are larger due to the artificial periodicity of the model system but we can qualitatively understand the role of the B subspace on the exciton singlet-triplet splitting. When the substrate adsorbate distance is reduced to 3.5\AA{}, including the substrate screening within the $S$ approximation reduces the singlet-triplet splitting by only 2 meV.

\subsection{Boron Nitride/Graphene Superlattice}

\begin{figure}[t]
\includegraphics[width=246.0pt]{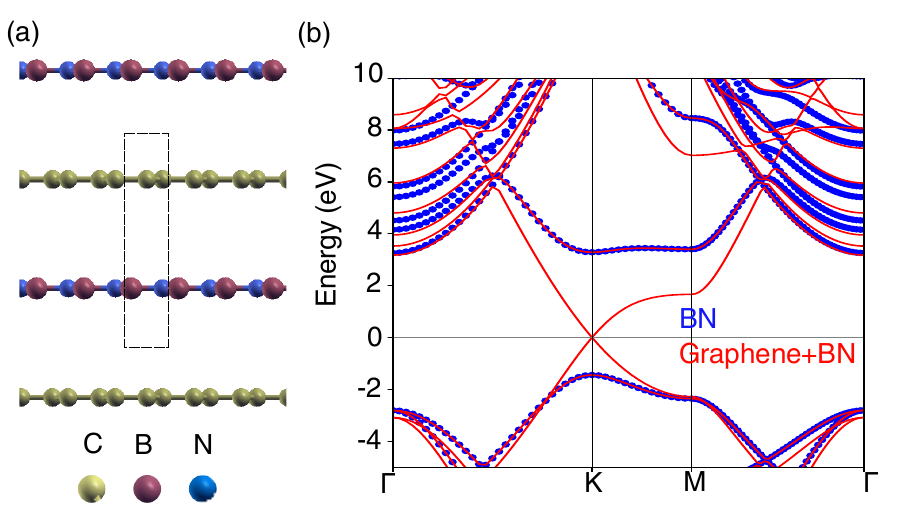}
\caption{\label{fig:gbn} (a) Illustration of supercell with a superlattice of alternating layers of graphene and monolayer h-BN. The dashed rectangle indicates the boundaries of the periodic supercell used in the calculation, which is periodic in all directions}. (b) Bandstructure at LDA level of the graphene and h-BN supercell (red lines) superimposed on the bandstructure of the isolated h-BN monolayer (blue circles).
\end{figure}

We next consider what happens when the external screening environment is metallic rather than semiconducting by constructing a bulk-like superlattice of alternating monolayers of h-BN and monolayers of graphene. Again, in order to make the calculation computationally tractable and reduce wave function hybridization, we construct an artificial supercell where the h-BN and graphene monolayers are separated by 5~\AA{} and the lattice of the h-BN is strained to match that of graphene (Fig.~\ref{fig:gbn}a). Figure ~\ref{fig:gbn}b shows the DFT bandstructure of the freestanding h-BN slab overlaid on the bandstructure of the h-BN/graphene superlattice to show that hybridization of the wavefunctions is minimal. We repeat the three kinds of calculations in Sec.~\ref{sec:mos2_bn} for the h-BN/graphene superlattice.

We find that similar to the case of h-BN encapsulated in \MoS2, graphene has a very small effect on the singlet-triplet splitting of excitons in h-BN. Inclusion of the graphene in the full BSE Hamiltonian reduces the singlet-triplet splitting of the lowest energy exciton in h-BN by 5~meV or about 5\%. Once again, the $S$ approximation gives the same result as the conventional treatment using a freestanding h-BN with the direct term screened.

\subsection{Analysis of Singlet-Triplet Splitting}

Here, we connect the effect of the substrate screening on the exciton singlet-triplet splitting to the character of the exciton states in the adsorbate in order to understand why the effect of the external screening is small. The screened-exchange contribution to the BSE kernel matrix elements in the $S$ approximation are given by
\begin{align}
    \langle v c \mathbf{k},\mathbf{Q} | K^{SX} | v' c' \mathbf{k'}, \mathbf{Q'}\rangle =
    \int d^3r d^3r'
    \psi^*_{v\mathbf{k}}(\mathbf{r})
    \psi_{c\mathbf{k+Q}}(\mathbf{r})
    \overline{W}(\mathbf{r-r'})
    \psi_{v'\mathbf{k}}(\mathbf{r'})
    \psi^*_{c'\mathbf{k'+Q'}}(\mathbf{r'}),
\end{align}
where $\overline{W}$ is the Coulomb interaction screened within the $B$ subspace of the BSE and evaluated within the TDA, as discussed in the previous sections.

The matrix elements can also be transformed into a Wannier basis
\begin{equation}
\label{eqn:wannier}
\begin{gathered}
\langle v\mathbf{R}_v; c\mathbf{R}_c | K^{SX} | v'\mathbf{R}_{v'}; c'\mathbf{R}_{c'}\rangle = \\
\int d^3r d^3r'
\mathcal{W}^*_{v}(\mathbf{r}-\mathbf{R}_v)
\mathcal{W}_{c}(\mathbf{r}-\mathbf{R}_c)
\overline{W}(\mathbf{r-r'})
\mathcal{W}_{v'}(\mathbf{r}'-\mathbf{R}_{v'})
\mathcal{W}^*_{c'}(\mathbf{r}'-\mathbf{R}_{c'})
\end{gathered}
\end{equation}

where $\mathcal{W}_n(\mathbf{r}-\mathbf{R})$ is a Wannier function for band $n$ centered at $\mathbf{R}$, and $v$ and $c$ label valence and conduction bands, respectively.

Since Wannier functions are well-localized in real space,the singlet-triplet splitting is dominated by matrix elements in Eq.~\ref{eqn:wannier} that correspond to having the electron and the hole in the same position, and so, for a two-band Wannier exciton, it is proportional to
\begin{align}
\Delta_{ST} \sim \langle v\mathbf{R}; c\mathbf{R} | K^{SX} | v\mathbf{R}; c\mathbf{R}\rangle =
\int d^3r d^3r'
\mathcal{W}^*_{v}(\mathbf{r})
\mathcal{W}_{c}(\mathbf{r})
\overline{W}(\mathbf{r-r'})
\mathcal{W}_{v}(\mathbf{r}')
\mathcal{W}^*_{c}(\mathbf{r}').
\end{align}

Because of the localization of the Wannier functions, the integral above is given by the screened Coulomb interaction with separations constrained to the dimension of the unit cell, a short length scale at which the screening due to the substrate $\overline{\epsilon}^{-1}(\mathbf{r},\mathbf{r}')$ is not effective. Hence, the screening due to the substrate will not lead to a modification in the singlet-triplet exciton energies, unless the separation between the substrate and the sample of interest is smaller than the typical size of the Wannier function. This is not the case in TMD monolayers, where the Wannier functions are smaller than the unit cell,~\cite{Sgiarovello2001,Li2017} and, more generally, we predict the singlet-triplet splitting in most quasi-2D systems will not be influenced by the screening due to the substrate. However, the substrate can still affect the singlet-triplet splitting in other systems, such as adsorbates that are metals or molecules, where the extent of the Wannier functions are larger than the substrate-adsorbate distance. This is consistent with previous calculations~\cite{Deilmann2019} that show that the exciton singlet-triplet splitting in a benzene molecule can be reduced by over 10\% when the molecule-substrate separation is smaller than the typical size of the Wannier or Boys function of the frontier orbitals in a benzene molecule ($\lesssim 5\mathrm{\AA}$).

\section{Summary}

In summary, we rederive the $S$ approximation using the L{\"o}wdin partitioning--obtaining a new expression with an additional factor of 1/2--and perform a systematic analysis of its accuracy by comparing the solution of the BSE with and without the $S$ approximation as a function of the size of subspace $A$ of the Hilbert space. We find that screening the exchange interaction in the BSE with a screening carefully designed to be consistent with the TDA can moderately speed up convergence with respect to subspace size in confined systems, such as molecules, where the exchange interaction couples a large number of occupied and unoccupied states in the same spatial region. We find that both metallic and semiconducting substrates have a negligible effect on singlet-triplet splitting in layered materials, where the wavefunctions of the system of interest and substrate do not significantly hybridize. In general, the screening from a substrate will not affect the exciton singlet-triplet splitting in the sample unless the extent of the maximally localized Wannier function for the single-particle states is larger than the adsorbate-substrate distance.

This work was supported by the Center for Computational Study of Excited State Phenomena in Energy Materials (C2SEPEM) at the Lawrence Berkeley National Laboratory, which is funded by the U.S. Department of Energy, Office of Science, Basic Energy Sciences, Materials Sciences and Engineering Division under Contract No. DE-AC02-05CH11231, as part of the Computational Materials Sciences Program, which provided theoretical formulation, algorithm and code development. DYQ acknowledges resources from a user project at the Molecular Foundry supported by the Office of Science, Office of Basic Energy Sciences, of the U.S. Department of Energy under the same contract number, which provided computational resources for the calculation on layered materials. This research used resources of the National Energy Research Scientific Computing Center (NERSC), a DOE Office of Science User Facility supported by the Office of Science of the U.S. Department of Energy under Contract No. DE-AC02-05CH11231, which provided computational resources for the calculation on benzene.

\bibliography{screened_exchange}
\bibliographystyle{apsrev4-1}

\end{document}